\documentclass{aipproc}

\usepackage{epsfig}

\def\lsim{\raise0.3ex\hbox{$<$\kern-0.75em\raise-1.1ex\hbox{$\sim$}}}
\def\gsim{\raise0.3ex\hbox{$>$\kern-0.75em\raise-1.1ex\hbox{$\sim$}}}

\newcommand\selectedlayoutstyle{6x9}
\layoutstyle\selectedlayoutstyle

\SetInternalRegister\hbadness{8000} 

\newcommand{\ie}{{\sl i.e.~\/}} 

\begin{document}

\title{Thermodynamics of 2 and 3 flavour QCD}


\author{F. Karsch}{
  address={Department of Physics, Bielefeld University, D-33615 Bielefeld, 
Germany},
  email={karsch@physik.uni-bielefeld.de}
}

\copyrightyear  {2001}

\begin{abstract}
We will discuss recent results on the thermodynamics of QCD in
the presence of light dynamical quark degrees of freedom. In
particular, we will concentrate on an analysis of the 
flavour and quark mass dependence of the QCD phase diagram, the 
equation of state and the transition temperature. Moreover, we 
present recent results on the heavy quark free energy.
\end{abstract}

\maketitle

\section{Introduction}

Lattice calculations have provided a rather detailed picture
of the thermodynamics of gluonic matter. We know that a phase 
transition exists, which separates a confining low temperature
phase (glueball gas) from a deconfined gluon gas phase at 
high temperature. The transition between these two phases
is first order, as originally predicted by Svetitsky and Yaffe
\cite{Svetitsky}. 
Although the low temperature phase only 
consists of rather heavy glueballs, the lightest of which has a mass of 
about 1.8~GeV \cite{glueball}, the phase transition temperature turns 
out to be rather small, $T_c = 0.637(5) \sqrt{\sigma} \simeq 270$~MeV.
This suggests that $T_c$ does not sensitively depend on the
mass of the lightest excitations in the confining phase. In fact,
it agrees quite well with values extracted from resonance gas
models based on the excitation spectrum of a gluonic 
string \cite{gluestring}.
Bulk thermodynamic observables like the 
energy density and pressure change rapidly at $T_c$ and
asymptotically approach the ideal gas limit. 
Their rapid rise signals the liberation of many new light degrees 
of freedom at $T_c$. However, even at temperatures a few times $T_c$ 
one still observes significant deviations from the asymptotic ideal gas
behaviour. This suggests that also in the high temperature phase
non-perturbative effects play an important role, which give rise
to large screening masses and quasi-particle excitations. 

Gluonic matter is described by an $SU(3)$ gauge theory, which is 
obtained as the infinite quark mass limit of QCD. In this limit 
the entire fermionic sector of QCD decouples and does not contribute 
to the thermodynamics; quarks only serve as static sources (quenched
QCD). This allows, for instance, a study of thermal modifications (screening) 
of the forces between external charges, which shows that the linear 
confining potential weakens with increasing temperature; the
string tension decreases and vanishes at $T_c$. 
   
Some of these basic aspects of gluonic thermodynamics clearly will change 
drastically in the presence of light dynamical quark degrees of 
freedom. The asymptotic high temperature limit for bulk thermodynamic
observables like the energy density or pressure 
rises with increasing number of light degrees of freedom.
Absolute confinement is lost in the presence of quarks with finite
mass already at low temperatures; the heavy quark free energy will 
no longer diverge when one tries to separate a static quark anti-quark pair.  
Moreover, the order of the QCD phase transition and even its very existence 
will crucially depend on the number of light degrees of freedom and
their masses.

During recent years these basic qualitative changes in the thermodynamics
of QCD, which result from the presence of light quark degrees of freedom,
have been observed in lattice calculations \cite{Karpisa,Eji00}. 
We will discuss here the current status of their quantitative analysis.
This will make clear that
unlike in the quenched sector of QCD we did not yet reach a similarly
detailed quantitative understanding of the relevant parameters
that control the critical behaviour of QCD with a realistic spectrum
of quark masses. However, thermodynamic calculations on the lattice
steadily improve. This partly is due to the rapid development of computer
technology. However equally important has been and still is the
development of improved discretization schemes, \ie improved actions.
This is of particular relevance for thermodynamic calculations which
are not only sensitive to the long distance physics at the phase
transition but also probe properties at short distances in calculations
of e.g. the energy density or the heavy quark potential \cite{improvement}. 

In the following we will discuss some recent results on the flavour and 
quark mass dependence of QCD thermodynamics. We will not go into any
details on the lattice formulation of QCD at finite temperature, which
have been discussed elsewhere \cite{Schladming}.

\section{The QCD phase diagram}

The quark mass and flavour dependence of the QCD phase transition
at finite temperature and vanishing baryon number density has been
studied extensively in lattice calculations. The basic qualitative
and quantitative features expected from universality arguments 
\cite{Svetitsky,Wilczek} on the 
one hand and phenomenological considerations on the other hand have been 
reproduced by these calculations. The transition is found to be first order 
in the limit of infinitely heavy quarks as well as in the chiral limit
of 3-flavour QCD. In the case of 2-flavour QCD the transition is found
to be continuous. The current status of the analysis of universal properties 
in the chiral limit, however, is not really satisfying \cite{Laermann}. 
The demonstration that for 2-flavour QCD the transition belongs to the 
universality class of 3-d, O(4) symmetric spin models still is ambiguous.

The regions of first order transitions are separated from a broad
crossover region by lines of second order phase transitions.
These lines are expected to belong to the 
universality class of the 3-d Ising model \cite{Gav94}, which is 
quite remarkable as the global $Z(2)$ symmetry which gets restored
at these transitions is not a symmetry of the QCD 
Lagrangian. This makes it obvious that the bare couplings 
appearing in the QCD Lagrangian cannot be the relevant scaling fields 
that control the critical behaviour at these lines of second order
transitions. 
It therefore also is not obvious a priori what is the correct 
order parameter for these transitions. The 
energy-like and magnetization-like operators of the effective Hamiltonian
that controls the critical behaviour at this critical point will
be linear combinations of the basic fields appearing in the QCD 
Lagrangian and the relevant temperature-like and ordering-field like
couplings will be linear combinations of the bare parameters of the
QCD Lagrangian, \ie the gauge coupling $\beta \equiv6/g^2$ and the 
quark masses $m_q$. 

Understanding the critical behaviour and the relevant scaling fields 
in the vicinity of the {\it chiral critical line} at small values
of the quark masses is, of course, interesting
in its own rights. However, it eventually will also be of importance
for a discussion of the physics in the vicinity of the critical endpoint
which is expected to exist in the temperature-density phase diagram
\cite{Rajagopal}. In an extended phase diagram, which also includes the
dependence on the chemical potential, these critical points lie in the
same critical surface of second order transitions.

\begin{figure}[t]
\caption{The cumulant $B_4$ versus the bare quark mass calculated in 
3-flavour QCD on different size lattices with a standard staggered
fermion action \cite{Kar01c}. 
}
\epsfig{file=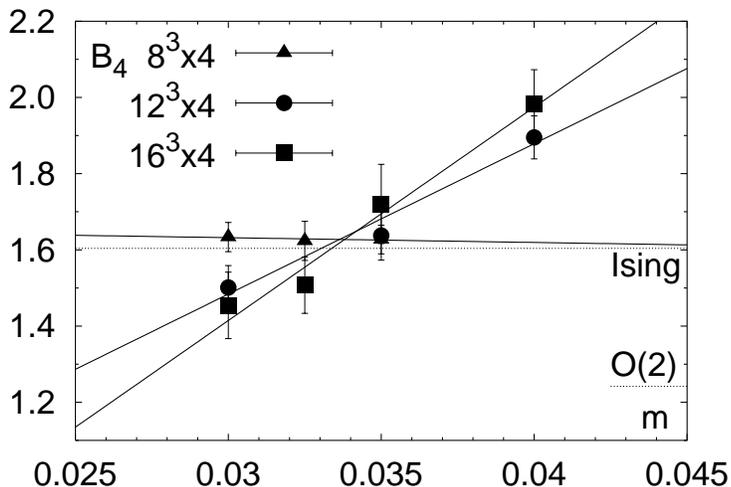,width=100mm}
\label{fig:Binder}
\end{figure}

The critical point separating the first order from the crossover
region has recently been studied in some detail for the case of
three degenerate quark mass flavours \cite{Kar01c}. Through an 
analysis of cumulants of the chiral condensate, 
$\langle \bar{\psi}\psi\rangle$, as well as joint probability 
distributions for the chiral condensate and
the gluonic part of the QCD action it could be shown that the 
transition indeed belongs to
the universality class of the 3-d Ising model and that the corresponding
order parameter can be constructed as a linear combination of the 
chiral condensate and the gluonic action. 
In Figure~\ref{fig:Binder} we show the result of a calculation of 
the fourth cumulant of the chiral condensate (Binder cumulant),
\begin{equation}
B_4 = {\langle (\bar{\psi}\psi)^4 \rangle
\over  \langle (\bar{\psi}\psi)^2\rangle^2}\quad .
\label{binder}
\end{equation}
The cumulant has been calculated for different values of the quark mass 
at the pseudo-critical couplings. Up to finite volume corrections 
the cumulants obtained on different size lattices will cross at 
the critical quark mass corresponding to the second order phase
transition point. The value of $B_4$ at this point is universal and
unambiguously identifies the transition as an Ising-like transition.

In order to judge whether QCD with a realistic spectrum of light 
u,d quarks and a heavier strange quark lies in the first order or
crossover region of the phase diagram it is necessary to determine
the location of the critical line quantitatively. The calculations 
performed so far with 3 degenerate quark masses and different fermion
actions \cite{Kar01c} show that the critical mass parameter,
e.g. the value of the pseudo-scalar meson mass
at the chiral critical point, is rather sensitive to cut-off effects.
While the calculation performed with the standard staggered fermion
action led to a critical value $m_{ps} \simeq 300$~MeV, calculations
with an improved staggered fermion action gave $m_{ps} \simeq 200$~MeV.
These calculations thus need further confirmation through studies
closer to the continuum limit.
Nonetheless, the current estimates consistently yield
rather small values for the pseudo-scalar meson mass at the
critical endpoint in 3-flavour QCD, which makes it quite 
unlikely that the physical point in the phase diagram, corresponding 
to a realistic quark mass spectrum with two light u,d, and a heavier 
strange quark, would lie in the first order region. This would also
be in agreement with an earlier estimate of the Columbia group \cite{Bro90}. 

\begin{figure}[htb]
\caption{The QCD phase diagram of 3-flavour QCD with degenerate
(u,d)-quark masses and a strange quark mass $m_s$. }
\vspace{-0.6truecm}
\epsfig{bbllx=59,bblly=175,bburx=564,bbury=514,
file=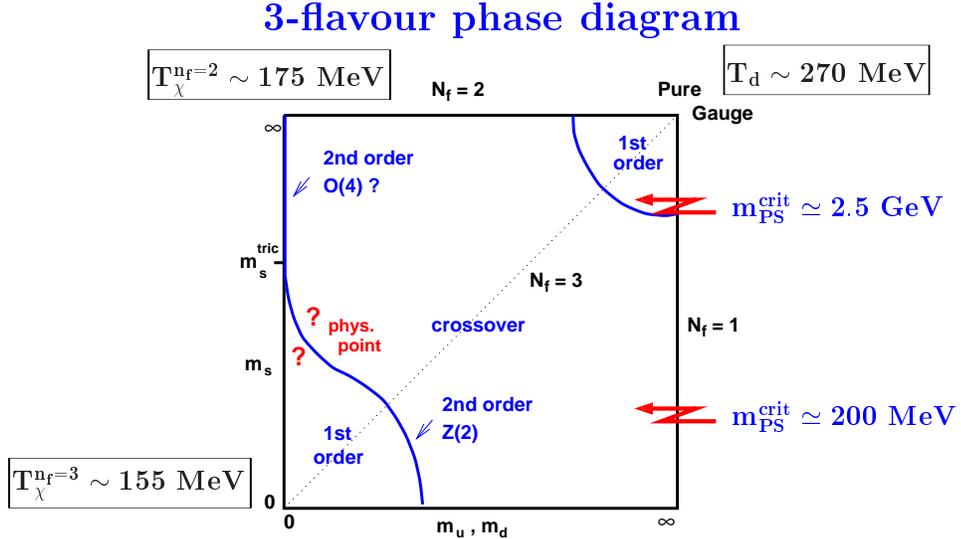,width=110mm}
\label{fig:phased}
\end{figure}

Our current understanding of the QCD phase diagram of 3-flavour QCD at 
vanishing baryon number density is summarized in Figure~\ref{fig:phased}.

\section{The Transition Temperature}

\begin{figure*}[t]
\caption{Transition temperatures in units of $m_V$ (left) and
in units of $\sqrt{\sigma}$ (right).  The left hand figure shows 
results for 2-flavour QCD obtained with different gauge and fermion actions.
The right hand figure shows the transition temperature in 2 (filled 
squares) and 3 (circles) flavour QCD obtained with improved staggered
fermions (p4-action). Also shown are results for 2-flavour QCD
obtained with the standard staggered fermion action (triangles).
The dashed band indicates the uncertainty on $T_c/\sqrt{\sigma}$ in the
quenched limit. The straight line is the fit given in Eq.~\ref{tcps}.
The vertical lines correspond to physical values of the hadron masses 
and a string tension of 425~MeV.
}
\hspace*{-0.2cm}\epsfig{file=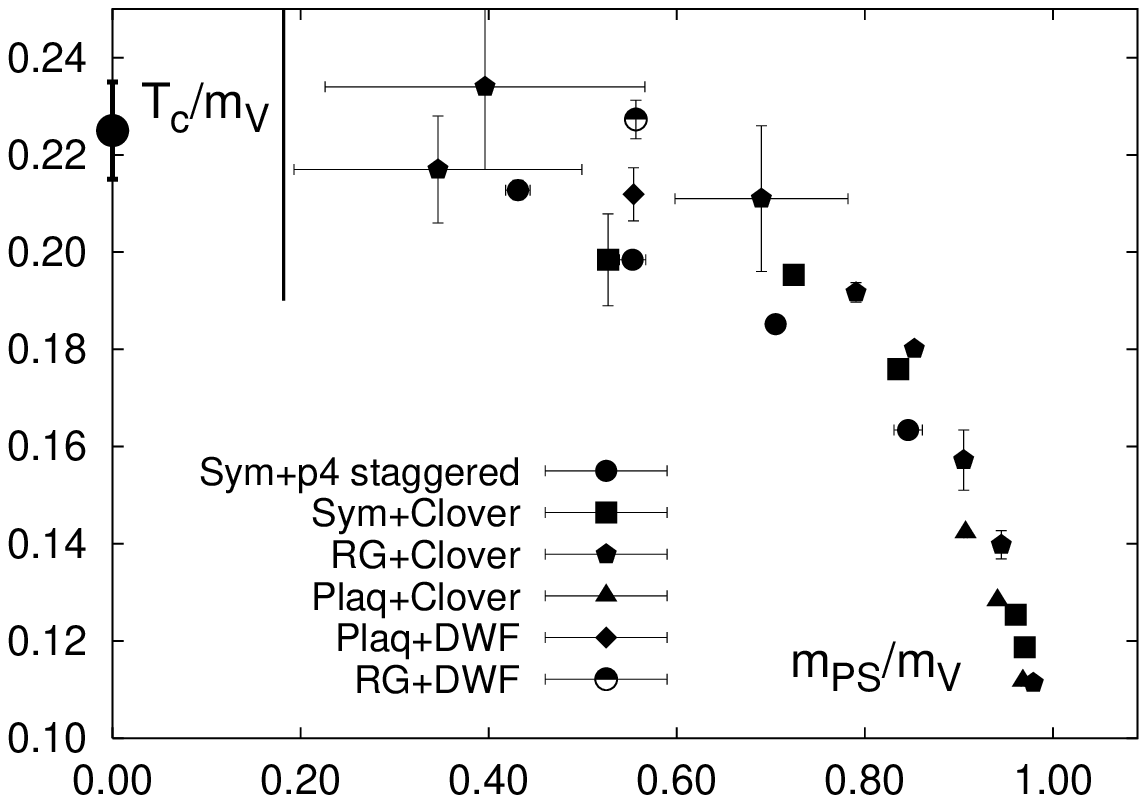,width=77mm} 
\hspace*{-0.1cm}\epsfig{file=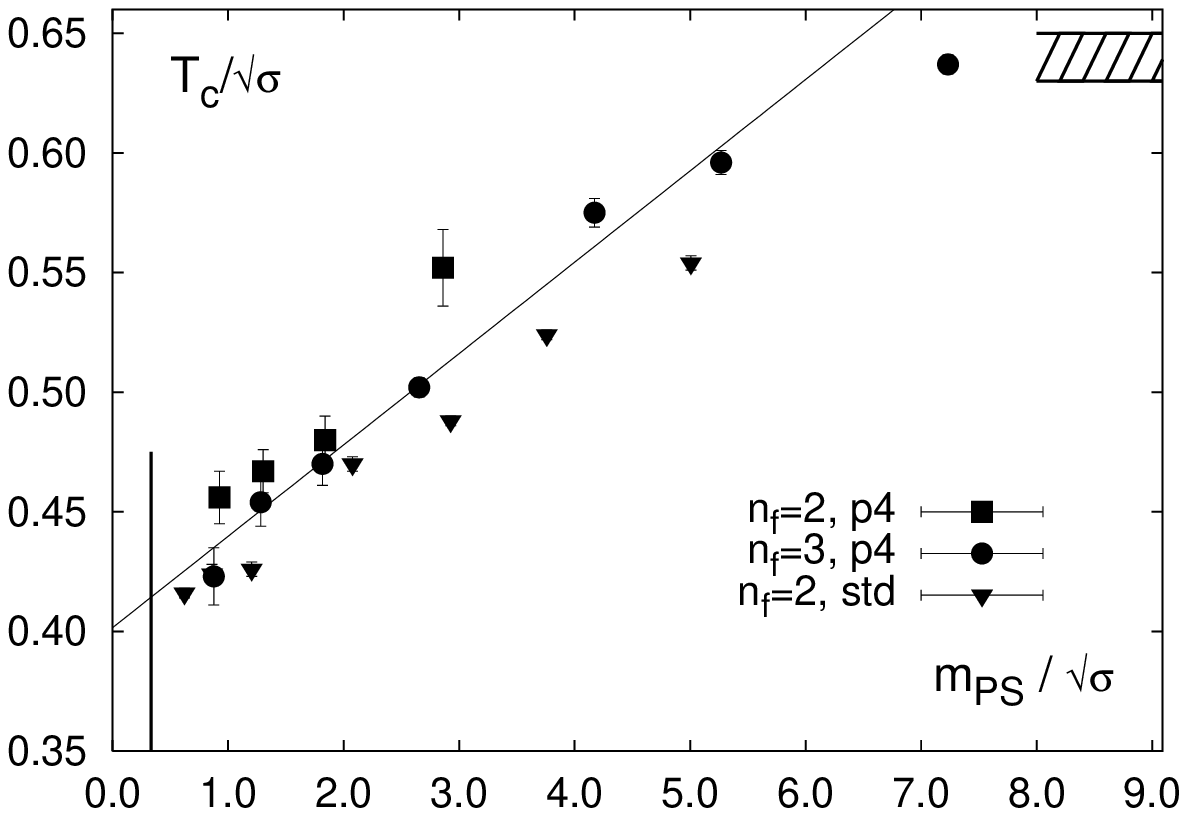,width=77mm}
\label{fig:tc}
\end{figure*}

It should be clear from the previous discussion of the phase 
diagram that for a large range of quark masses the transition to the
high temperature plasma phase is not a phase transition, \ie the
transition does not correspond to singularities in the QCD partition
function. Nonetheless, also for these quark masses the transition
occurs in a narrow temperature interval and 
thus is well localized. This transition region is characterized by  
peaks in susceptibilities, e.g. the chiral susceptibility $\chi_m$ or
the Polyakov-loop susceptibility $\chi_L$,
\begin{equation}
\chi_{m} = {\partial \over \partial m_q} \langle \bar{\psi}\psi
\rangle
\quad, \quad
\chi_L = N_\sigma^3\; \biggl(\langle L^2 \rangle - \langle L \rangle^2\biggr)~~
\quad .
\label{sus}
\end{equation}
Here 
\begin{equation}
L={1\over N_\sigma^{3}}\;\sum_{\vec{x}} {\rm Tr}\; L_{\vec{x}}
\label{poly}
\end{equation}
denotes the spatial average over Polyakov-loops\footnote{The 
Polyakov-loop is a purely time-like Wilson loop which is
closed due to the periodicity of the lattice in temporal direction.
Its definition as well as further details on the lattice formulation
of QCD thermodynamics may, for instance, be found in \cite{Schladming}.}, 
$L_{\vec{x}}$, defined at the spatial
sites $\vec{x}$ of a lattice of size $N_\sigma^3 \times N_\tau$.
A calculation of the
pseudo-critical temperature, $T_c = 1/N_\tau \; a(\beta_{pc})$ still
requires the determination of the lattice spacing $a(\beta_{pc})$
at the pseudo-critical couplings $\beta_{pc}$ which in turn are
defined through the location of the susceptibility peaks.
The lattice spacing can be determined through the calculation
of an independent physical observable. In order to quote $T_c$ in
physical units, \ie MeV, we need a physical observable
which does not crucially depend on the values of the quark masses.
For physical values of the 
quark masses it would, of course, be most appropriate to use a hadron
mass, e.g. the rho-meson mass, to set the physical scale for $T_c$.
The hadron masses, however, are themselves strongly dependent on the
quark mass values; their masses diverge in the infinite quark
mass limit. Nonetheless, we can assign a physical value to
the transition temperature in this limit. In quenched QCD the natural 
scale for $T_c$ is the square root of the string tension, $\sqrt{\sigma}$. 
In fact, the string tension as well as quenched hadron 
masses\footnote{In the calculation of {\it quenched hadron masses}
the contribution of dynamical light sea quarks is suppressed. Only
the valence quark contributions and interactions with the gluonic
vacuum are taken into account.} seem to
describe the experimentally known QCD spectrum reasonably well
already in the infinite quark mass limit. These quantities thus seem to
have a weak quark mass dependence and are suitable to set the scale
for $T_c$ at arbitrary values of $m_q$. This situation is illustrated
in Figure~\ref{fig:tc}.

On the left hand side of Figure~\ref{fig:tc} we show the transition
temperature in units of the vector meson mass $m_V$ versus the ratio
$m_{ps} / m_V$, which in the chiral limit is proportional to the square
root of the quark mass. Here the drop in $T_c/ m_V$ observed with
increasing quark mass mainly is due to 
the quark mass dependence of $m_V$ and does not reflect the quark
mass dependence of $T_c$. This figure shows, however, that calculations
based on different discretization schemes for the gauge and fermion
actions do yield consistent results for $T_c$ and its dependence on
$m_q$.
In the chiral limit one finds for the critical temperature in 
2 and 3-flavour QCD

\begin{eqnarray}
\underline{\rm 2-flavour~ QCD:} &&
T_c  = \cases{(171\pm 4)\; {\rm MeV}, &  clover-improved
Wilson fermions \cite{Ali01} \cr
(173\pm 8)\; {\rm MeV}, & improved staggered fermions  \cite{Kar00a}}
\nonumber \\
\underline{\rm 3-flavour~ QCD:} &&
T_c  = \; \; \; \;
(154\pm 8)\; {\rm MeV}, \; \; \;  {\rm improved~ staggered~ fermions~}
[15]
\nonumber
\end{eqnarray}
\vspace{0.1cm}

Here $m_\rho$ has been used to set the scale for $T_c$. We note that 
all the results presented in Figure~\ref{fig:tc} have been obtained on 
lattices with a rather small temporal extent ($N_\tau =4$). The lattice 
spacing is therefore still
quite large in the vicinity of $T_c$, \ie $a\simeq 0.3$~fm. 
Moreover, there are uncertainties involved in the ansatz used to
extrapolate to the chiral limit. We estimate that the systematic
error on the value of $T_c /m_\rho$ still is of similar magnitude as
the purely statistical error quoted above. The critical temperature
thus is at present only known with an error of about 10\%.
 
On the right hand side of Figure~\ref{fig:tc} we give $T_c$ in units
of $\sqrt{\sigma}$. This reflects the expected behaviour; with 
increasing values of the quark mass the hadrons become heavier and it 
becomes more difficult to create a dense hadronic system that could
undergo a transition to a quark-gluon plasma phase; the transition 
temperature thus increases. 
It is, however, striking that the quark mass dependence of $T_c$
is so weak.
The straight line shown in Figure~\ref{fig:tc} is a fit to the 3-flavour
data, which gave
\begin{equation}
(T_c / \sqrt{\sigma} )_x = (T_c / \sqrt{\sigma} )_0 +
0.04(1)\; x \quad {\rm with} \quad x =m_{ps}/ \sqrt{\sigma} \quad .
\label{tcps}
\end{equation} 
This suggests that also heavier resonances, which are little affected
by changes of the quark mass, play an important role for building up
the critical density needed to trigger the QCD phase transition.
Moreover, we note that for pseudo-scalar masses larger than about 
$6\sqrt{\sigma}\simeq 2.5$~GeV the transition temperature agrees with the 
value found in the 
pure gauge theory, $T_c \simeq 0.637 \sqrt{\sigma} \simeq 270$~MeV.
In this heavy quark mass regime all the hadronic states are heavier 
than typical glueball masses and thus decouple from the thermodynamics, 
which only is controlled by gluonic degrees of freedom. In this
heavy mass regime the transition also changes again from a crossover 
to a first order phase transition. 

\section{The QCD equation of state}

The temperature dependence of bulk thermodynamic observables like
the energy density ($\epsilon$) and pressure ($p$) has been analyzed 
in detail in the pure gauge sector. These calculations show that 
$\epsilon/T^4$ as well as $p/T^4$ rapidly increase above $T_c$. However,
even at $T\; \simeq 4\;  T_c$ the Stefan-Boltzmann limit is not yet 
reached \cite{Boyd}. Deviations stay at the level of 15\%, which is too
much to be understood in terms of perturbative corrections 
\cite{perturbation}. 
Even at these high temperatures non-perturbative effects seem to play an
important role for the thermodynamics. This finds further support from 
studies of the heavy quark free energy \cite{Kaczmarek}, the spatial string 
tension \cite{spatial} as well as the gluon propagator in fixed gauges 
\cite{propagator}. The analysis of the temperature dependence of these
observables also suggests that the temperature dependent
running coupling constant remains large in the plasma phase and 
electric and magnetic screening masses differ significantly
from perturbative results even at temperatures which are
several orders of magnitude larger than $T_c$ \cite{Kajantie}. 
In view of this it may even be surprising that
models with weakly interacting quasi-particles \cite{quasi} as well as
calculations based on self-consistent resummation schemes \cite{hight}
do provide a quite satisfactory description of bulk thermodynamics down
to temperatures a few times $T_c$.  

At least at high temperature, where the energy density and pressure
are expected to approach the free gas limit, the QCD equation of state 
will strongly depend on the number of light partonic degrees
of freedom. Already for two massless quark flavours the Stefan-Boltzmann 
constant increases by more than a factor two relative to
the pure gauge theory. For $n_f$-flavour QCD one has,
\begin{equation}
{\epsilon_{SB} \over T^4} = {3p_{SB} \over T^4} =
\biggl( 16 + {21 \over 2} n_f\biggr) {\pi^2 \over 30} \quad .
\label{esb}
\end{equation}
As the influence of non-zero quark masses on bulk thermodynamic
observables will be exponentially suppressed at high temperature, 
Eq.~\ref{esb} also gives the dominant high temperature behaviour for massive
quark even when the masses are of the order of the (critical) temperature. 
In fact, this has been observed in lattice calculations with fairly large 
quark masses \cite{Ali01b}.

\begin{figure}[t]
\caption{The pressure in QCD with $n_f = 0,~2$ and 3 light quarks as
well as two light and a heavier (strange) quark. For $n_f \ne 0$
calculations have been performed on a $N_\tau=4$ lattice using
improved gauge and staggered fermion actions \cite{Kar00a}. In the case of 
the SU(3)
pure gauge theory the continuum extrapolated result is shown \cite{Boyd}. 
Arrows indicate the ideal gas pressure $p_{SB}$ as given in Eq.~\ref{esb}.
}
\hspace*{-0.2cm}\epsfig{file=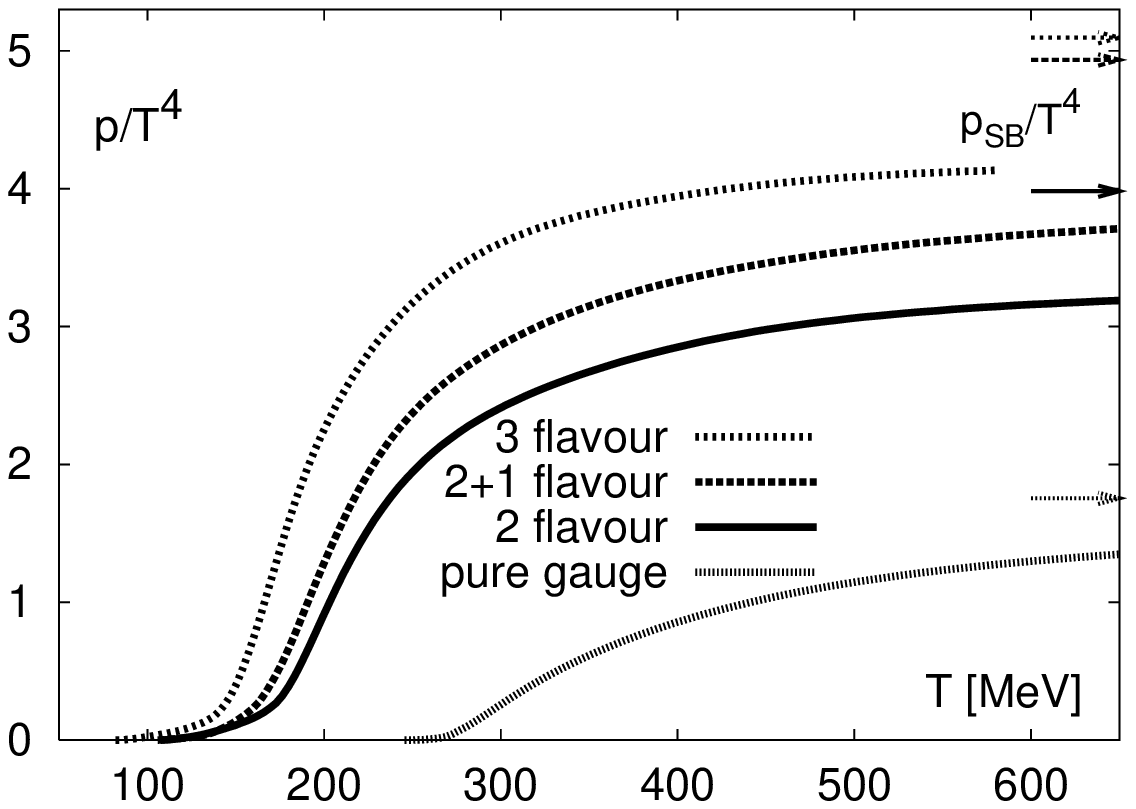,width=78mm} 
\hspace*{-0.3cm}\epsfig{file=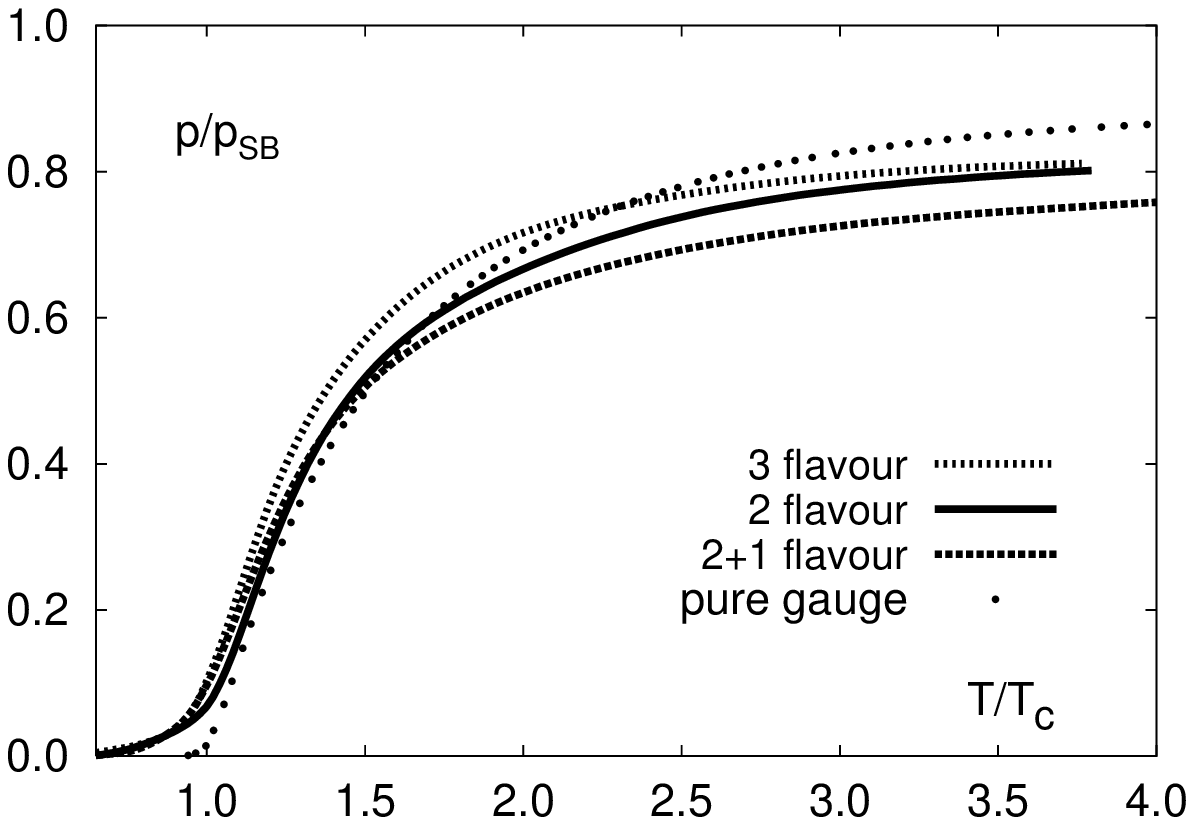,width=78mm}
\label{fig:eos}
\end{figure}

The overall pattern of the temperature dependence of bulk thermodynamic
observables in QCD with 2 and 3 light quark flavours is very similar to 
the case of the pure gauge theory. With increasing number of light
degrees of freedom the critical temperature shifts to smaller values  
and the asymptotic high temperature limit for $p/T^4$ becomes larger. 
This is seen
in the left hand part of Figure~\ref{fig:eos}. However, after rescaling
the thermodynamic observables with the corresponding Stefan-Boltzmann
constants they look quite alike in units of $T/T_c$. Of course, they
still differ in details. In particular, it is apparent from the right 
hand part of Figure~\ref{fig:eos} that at $T_c$ the rescaled pressure of
QCD with light quarks is significantly larger than in the pure gauge
theory. This also is the case for the energy density which is found
to be $\epsilon_c /T_c^4 \simeq 6$ in QCD with light quarks 
\cite{Ali01b,energy}, while it
is only $\epsilon_c /T_c^4 \simeq 1$ in the SU(3) gauge theory\footnote{The
energy density is discontinuous in the case of the pure gauge theory.
The latent heat is found to be $\Delta\epsilon /T_c^4 = 1.40(9)$ \cite{Bei97}
with $\epsilon /T_c^4 \simeq 2$ in the high temperature phase.}.
Much of this factor 6 difference in $\epsilon_c /T_c^4$, however, seems to
arise from the difference in $T_c$ between QCD with light quarks and 
the purely gluonic theory. The critical energy densities turn out 
to be quite similar. 
Unfortunately, the current error on $T_c$, which is about 10\%, 
amplifies in the calculation of the energy density, which makes
$\epsilon_c$ still badly determined in lattice calculations,
\begin{equation}
\epsilon_c \; =\; (0.3 - 1.3)\; {\rm GeV  / fm}^3 \quad .
\label{ec}
\end{equation}

\section{The heavy quark free energy}

The confining properties of the thermal heat bath generated by quarks and
gluons can be probed by analyzing the response of the medium to the insertion 
of static sources. 
Static quark sources are described by the Polyakov-loop $L_{\vec{x}}$.
In the pure gauge limit the expectation value of
its spatial average, $\langle L \rangle$ (Eq.~\ref{poly}), is an order 
parameter for the deconfinement transition,  
\begin{equation}
\langle L \rangle \; \cases{=\; 0\; ,& $T\; < \; T_c$ \cr
> \; 0\; ,& $T\; > \; T_c$ } \quad .
\label{order}
\end{equation}
This reflects the long distance behaviour of the correlation function
for static quark anti-quark sources,
\begin{equation}
{\rm e}^{-F(r,T)/T} \; = \; 
\langle {\rm Tr} L_0 \; {\rm Tr} L^{\dagger}_{\vec{x}} \rangle 
\quad , \quad 
r\equiv |\vec{x}| \quad ,
\label{polcor}
\end{equation}
which approaches the cluster value $|\langle L \rangle|^2$ in the 
limit of infinite separation between the quark anti-quark pair.
In the confined phase this correlation function vanishes for
$r\rightarrow\infty$ which signals that the free energy needed to 
separate the two sources is infinite.
This is no longer the case if dynamical quarks with a finite mass are 
contributing to the thermal heat bath. The static sources can then be 
screened by quarks and anti-quarks present in the heat bath and even in 
the zero temperature limit this becomes possible through the generation 
of $q\bar{q}$-pairs from the vacuum. The free energy needed to separate 
the static sources thus will stay finite at all temperatures. 

\begin{figure}[t]
\caption{The heavy quark free energy in units of the square root of
the string tension versus the quark anti-quark separation (left)
and the estimate for the dissociation energy defined in Eq.~\ref{diss}  
(right). The band shows the normalized Cornell potential
$V(r) = -\alpha/r + \sigma \; r$ with $\alpha = 0.25\pm0.05$.
The lattice data have been normalized to this potential at the 
shortest distance available, \ie at $rT=1/4$.
}
\hspace*{-0.2cm}\epsfig{file=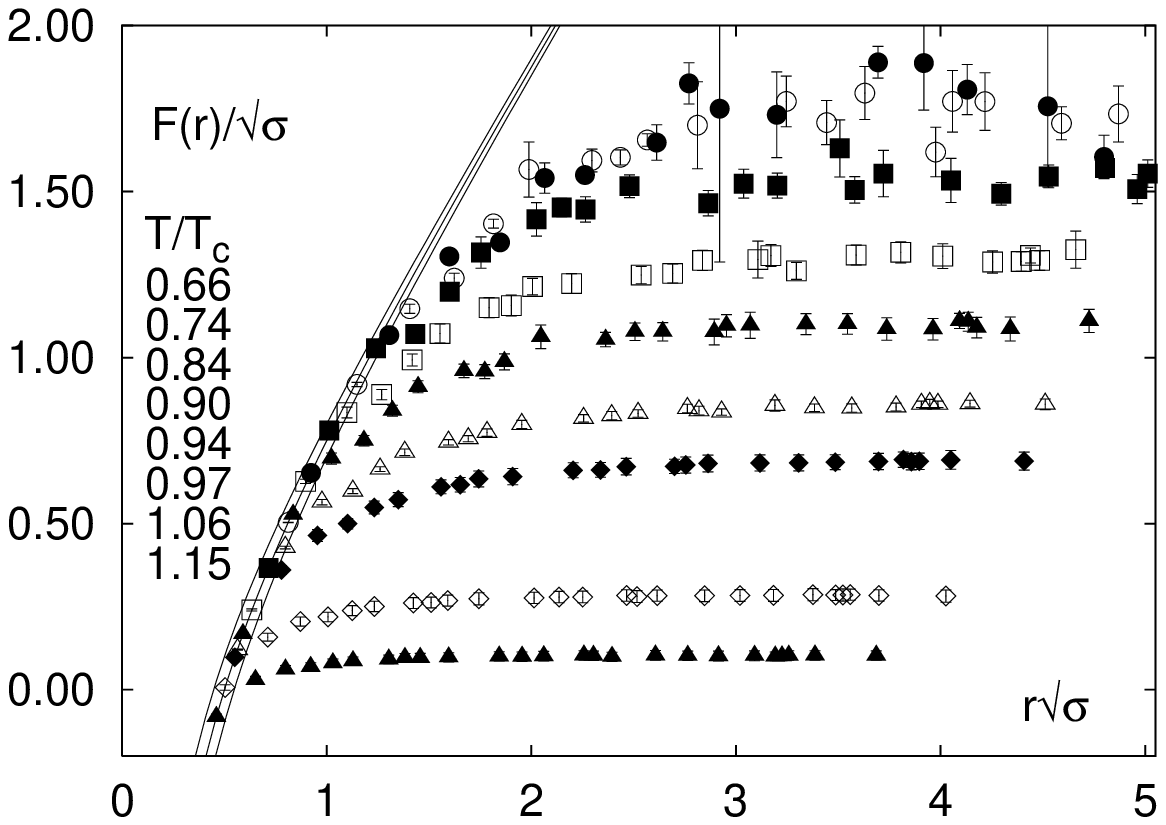,width=77mm}
\hspace*{-0.2cm}\epsfig{file=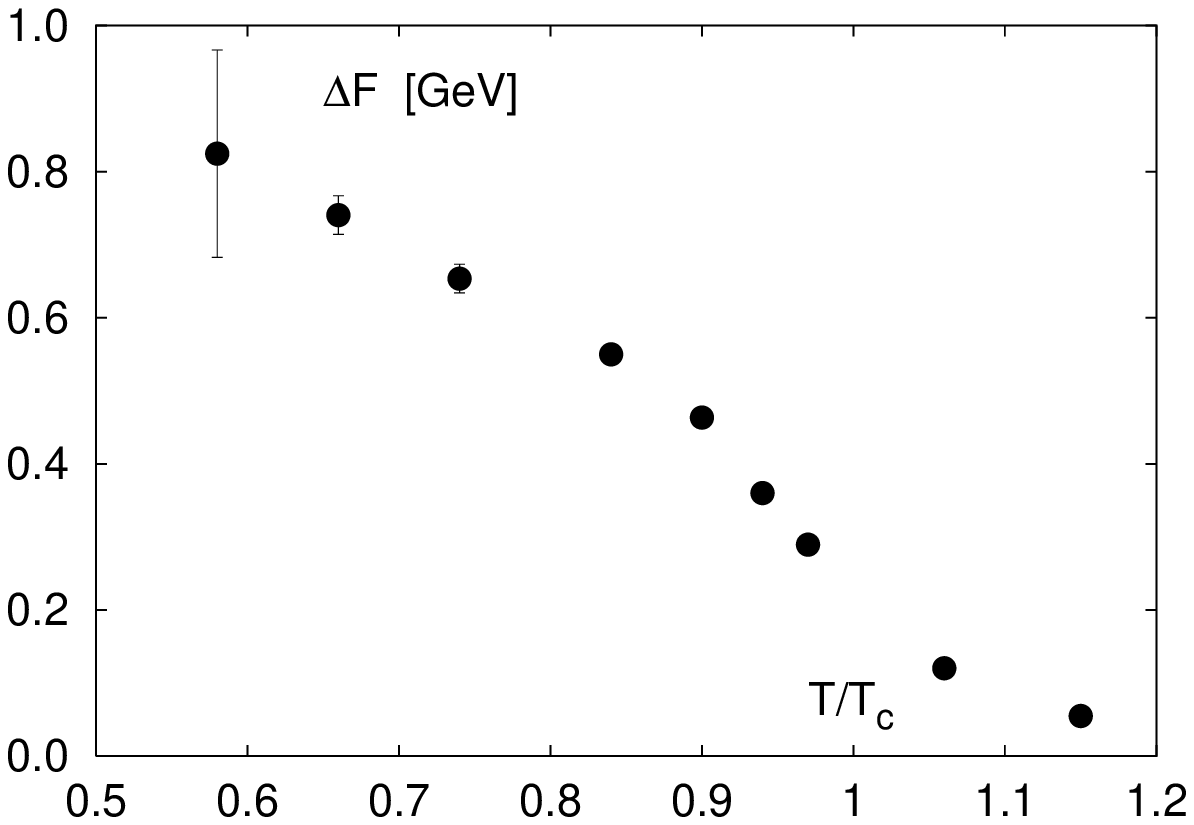,width=77mm}
\label{fig:potential}
\end{figure}

In Figure~\ref{fig:potential} we show the free energy of a static quark 
anti-quark pair calculated in 3-flavour QCD at various temperatures
and for different separation of the static sources. 
At low temperatures, $T\lsim 0.7\; T_c$, the heavy quark free energy
coincides with the confining Cornell-type potential,
$V(r) = -\alpha/r + \sigma \; r$ with $\alpha = 0.25\pm0.05$, up 
to distances $r \simeq 1.5/\sqrt{\sigma} \simeq 0.7$~fm. With 
increasing temperature $F(r,T)$, however, gets screened earlier
and at $T_c$ it starts to deviate from the zero temperature potential 
already at distances $r\simeq 0.3$~fm. Moreover, it becomes much 
easier to separate the heavy quark sources. As an estimate for
the dissociation energy we show in the right hand part of 
Figure~\ref{fig:potential} the difference in free energy of a 
$q\bar{q}$-pair at infinite separation and a $q\bar{q}$-pair
at distance $r_q=\sqrt{\alpha /\sigma}$, 
\begin{equation}
\Delta F \equiv \lim_{r\rightarrow \infty}F(r,T) -
F(\sqrt{\alpha / \sigma})\quad .
\label{diss}
\end{equation}

The rapid decrease of $\Delta F$ close to $T_c$ clearly will have 
consequences for the formation and existence of heavy quark bound 
states not only in the high temperature phase but also 
in the vicinity of $T_c$. Already at $T\simeq 0.9 \; T_c$
the free energy difference for a heavy quark pair separated by a distance
similar to the $J/\psi$ radius ($r_\psi \sim 0.2$~fm) and a $q\bar{q}$-pair
at infinite separation is only 500~MeV, which is compatible with the average
thermal energy of a gluon ($\sim 3\; T_c$). The $c\bar{c}$-bound states are 
thus expected to dissolve already close to $T_c$, maybe even below $T_c$
\cite{Satz01}.

\begin{theacknowledgments}
The work has been supported by the TMR network ERBFMRX-CT-970122 and by
the DFG under grant FOR 339/1-2.

\end{theacknowledgments}

\vspace*{0.5cm}


\end{document}